\documentclass[aps,prb,twocolumn,groupaddress]{revtex4}
\usepackage{epsfig}
\usepackage[usenames]{color}
\DeclareMathAlphabet{\mathitb}{OT1}{cmr}{bx}{sl}

\begin{document}

\renewcommand{\thefootnote}{\fnsymbol{footnote}}

\title{Back-action Driven Electron Spin Singlet-Triplet Excitation in a Single Quantum Dot}

\author{Gang Cao}
\author{Ming Xiao}
\email{maaxiao@ustc.edu.cn}
\author{HaiOu Li}
\author{Cheng Zhou}
\author{RuNan Shang}
\author{Tao Tu}
\author{GuangCan Guo}
\author{GuoPing Guo}
\email{gpguo@ustc.edu.cn}
\affiliation{Key Laboratory of Quantum Information, Chinese Academy of Sciences, University of Science and Technology of China, Hefei 230026, People’s Republic of China}
\author{HongWen Jiang}
\affiliation{Department of Physics and Astronomy, University of California at Los Angeles, 405 Hilgard Avenue, Los Angeles, CA 90095, USA}

\date{\today}

\begin{abstract}
In a single quantum dot (QD), the electrons were driven out of thermal equilibrium by the back-action from a nearby quantum point contact (QPC). We found the driving to energy excited states can be probed with the random telegraph signal (RTS) statistics, when the excited states relax slowly compared with RTS tunneling rate. We studied the last few electrons, and found back-action driven spin singlet-triplet (S-T) excitation for and only for all the even number of electrons. We developed a phenomenological model to quantitatively characterize the spin S-T excitation rate, which enabled us to evaluate the influence of back-action on spin S-T based qubit operations. 
\end{abstract}

\maketitle

Individual electron charges or spins in semiconductor quantum dots (QDs) are prospective for implementing solid-state quantum computers \cite{Charge-Qubits, Spin-Qubits}. Especially, the spin singlet-triplet (S-T) states have been widely utilized to realize single and double qubits \cite{Spin-Qubits}. In nearly all these demonstrations, a quantum point contact (QPC) is needed to read out the qubit states. However,  it is also speculated that the inevitable QPC back-action can cause the qubit states to relax and dephase \cite{Backaction-Relax-Dephase}. Experimentally, it lacks a quantitative study of the effects of back-action on the evolution of spin S-T states, which will be important to achieve high fidelity quantum computation.

In this work, we observed back-action induced excitation from spin singlet to triplet states and quantitatively determined the excitation rate. The back-action strength is tunable through the QD-QPC coupling or QPC bias. Under strong back-action condition, the spin singlet electron is driven out of the QD or up to the triplet states. Due to its sensitivity to the QD energy spectroscopy, the QPC real-time charge counting statistics serves as a probe to the spin S-T excitations.  By solving density matrix rate equations, we can quantitatively measure the excitation rate, which will enable us to evaluate the effect of back-action on spin S-T based qubits.

\begin{figure}[t]
\begin{center}
\epsfig{file=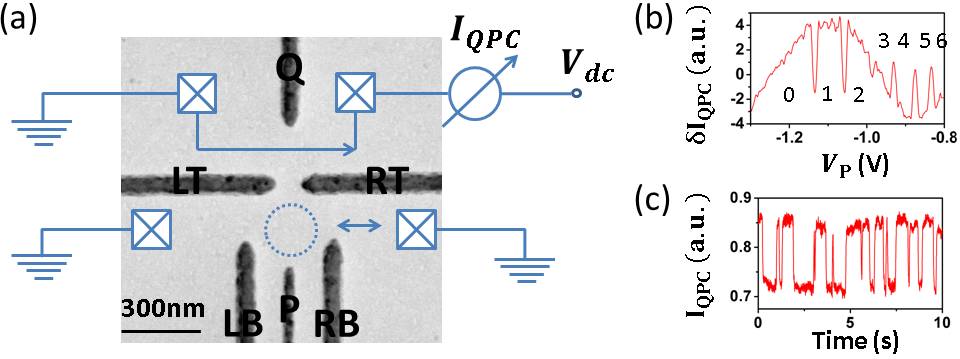, width=1\linewidth, angle=0}
\end{center}
\vspace{-7.5mm}
\caption{(a) A diagram showing the geometry of our QD-QPC structure and the measurement set-up.  (b) Numerical derivative of $I_{QPC}$ when $V_{P}$ is swept to control the QD electron number from 6 to 0. (c) A typical trace of RTS when $V_{P}=-1.11V$, corresponding to the $0e \leftrightarrow 1e$ tunneling.}
\label{Figure1}
\end{figure}

A single QD with a QPC on side was fabricated in a GaAs/AlGaAs heterostructure, as shown in Fig. \ref{Figure1} (a). The left barrier of the QD is closed and the electrons only tunnel through the right barrier (tunneling rate controlled by gates RT and RB). A tiny gap between gates LT and RT, which opening is modulated by voltage $V_{T} \equiv V_{LT}=V_{RT}$, was found to control the QPC back-action to the QD \cite{Ming-Submitted-RTS-Backaction}. All the ohmic contacts are grounded except that one side of the QPC is applied with a small dc voltage $V_{dc}$ to measure the current $I_{QPC}$.  Fig. \ref{Figure1} (b) shows the numerical derivative of $I_{QPC}$, with each sharp peak representing the tunneling of an individual electron. In this work we studied the last six electrons. Fig. \ref{Figure1} (c) shows a random telegraph signal (RTS) trace, i.e., the real-time electron tunneling, for $0e \leftrightarrow 1e$. 

\begin{figure}[t]
\begin{center}
\epsfig{file=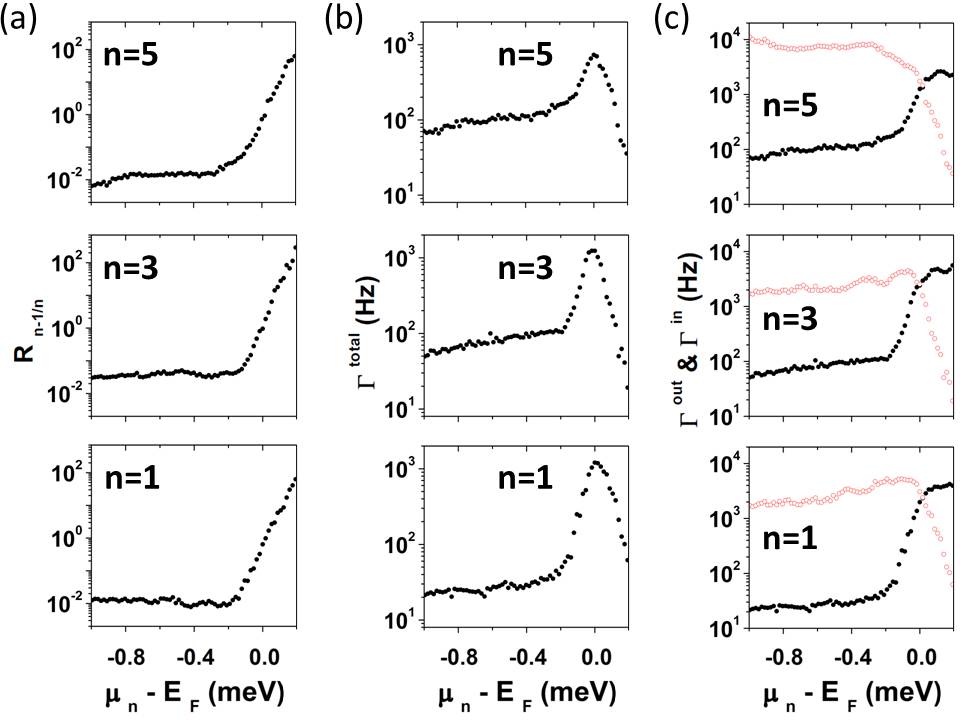, width=1\linewidth, angle=0}
\end{center}
\vspace{-7.5mm}
\caption{RTS statistics of $(n-1)e \leftrightarrow ne$ transition for odd electron numbers $n=5$, $3$, and $1$. Here $V_{T}=-1.2 V$ and $V_{dc}= 1mV$. (a) $R_{n-1/n}$. (b) $\Gamma^{total}$. (c) Black closed circles are $\Gamma^{out}$, and red open circles are $\Gamma^{in}$.}
\label{Figure2}
\end{figure}

The characterization of back-action strength was reported in an earlier work \cite{Ming-Submitted-RTS-Backaction}. The back-action, controlled by the QPC bias voltage or the QD-QPC opening,  contributes an extra tunneling out rate that can be measured using RTS statistics. In this paper, we found that the RTS statistics showed distinctive features for the odd and even number of electrons. We recorded the RTS statistics for all the last 6 electrons. First let's look at Fig. \ref{Figure2} for the odd numbers, $n=5$, $3$, and $1$. They just showed the back-action induced saturation effect as we explained earlier \cite{Ming-Submitted-RTS-Backaction}. When $\mu_{n} << E_{F}$, the $n^{th}$ electron is supposed to be trapped in the dot and the electron occupancy ratio $R_{n-1/n}$ should exponentially drop to zero. Instead, we see that $R_{n-1/n}$ saturates on the $\mu_{n} << E_{F}$ side because the $n^{th}$ electron can absorb the phonons emitted by QPC and escape the dot. Same effect is seen for the total tunneling rate $\Gamma^{total}$ and tunneling out rate $\Gamma^{out}$. The back-action induced tunneling out rate $\Lambda^{out}$ is found to be invariant within a certain cut-off energy. On the other hand there is no obvious saturation effect on the $\mu_{n} >> E_{F}$ side, which indicates that the back-action induced tunneling in rate $\Lambda^{in}$ is negligible since the 2-dimensional electron gas (2DEG) is a thermal equilibrium reservoir.  

\begin{figure}[t]
\begin{center}
\epsfig{file=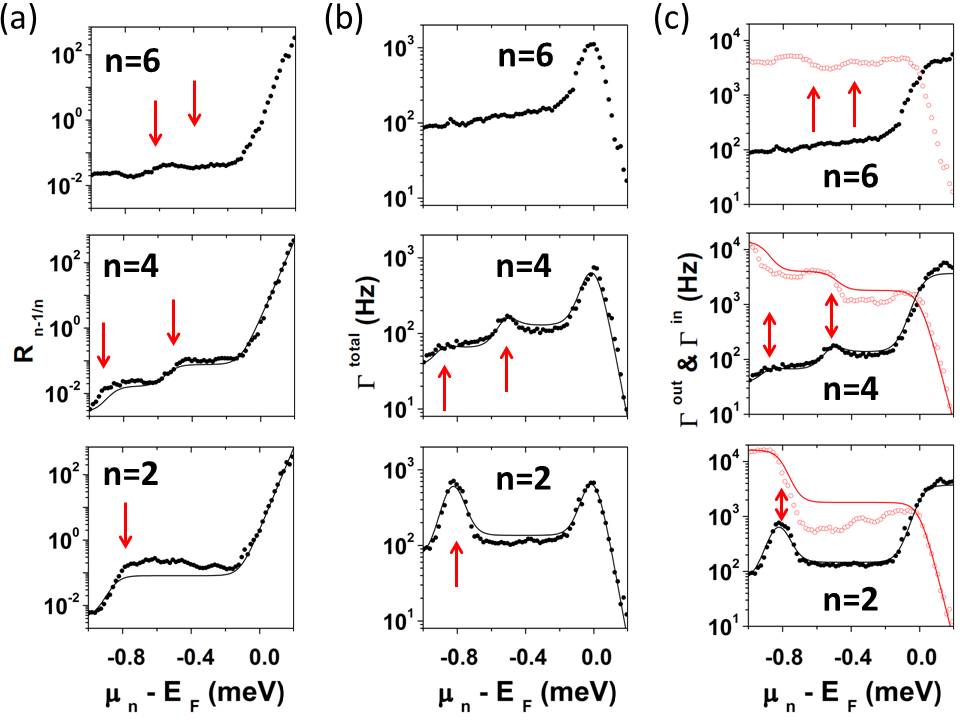, width=1\linewidth, angle=0}
\end{center}
\vspace{-7.5mm}
\caption{RTS statistics of $(n-1)e \leftrightarrow ne$ transition for even electron numbers $n=6$, $4$, and $2$. Here $V_{T}=-1.2 V$ and $V_{dc}= 1mV$. (a) $R_{n-1/n}$. (b) $\Gamma^{total}$. (c) Black closed circles are $\Gamma^{out}$, and red open circles are $\Gamma^{in}$. The red arrows point to the additional features. The solid lines are the simulation considering excited states.}
\label{Figure3}
\end{figure}

As a comparison, the even electron numbers showed dramatic additional features except this saturation effect, as indicated by the red arrows in Fig. \ref{Figure3}. For instance, for $n=2$, $R_{1/2}$ shows an extra elevated plateau extending to the point $\mu_{n} - E_{F}=-0.80meV$. Except the major peak at the balance point $\mu_{n} - E_{F}=0$, $\Gamma^{total}$ shows an extra peak also at $-0.80meV$. The extra feature is also seen in $\Gamma^{out}$ and $\Gamma^{in}$. For $n=4$, a pronounced extra feature is observed at $-0.50meV$. Another minor additional feature occurs at $-0.90meV$. For $n=6$, there are two likely extra features at $-0.38meV$ and $-0.62meV$, although almost buried in noise. 

These additional features apparently represent some excited states that the QD electrons hit into for some reason \cite{Hirayama-Backaction-RTS}. The finding that these excited states only appear for even number of electrons makes us to rationalize that they are the spin S-T splittings. A number of reasons justify our speculation. First of all, orbital excited states have much faster relaxation rate ($T_{1}$ in the $10 ns$ range in GaAs QDs \cite{Petta-Charge-Qubit-T1}) compared with the RTS tunneling rate (about $1 kHz$ at the balance point) and we believe they can hardly be resolved in this experiment. Spin excited states, on the other hand, have slow relaxation rate ($T_{1}$ approaches $ms$ for S-T \cite{Kouwenhoven-Spin-ST-T1}) comparable to the RTS tunneling rate here and should be readily resolved. Second, an even number of electrons are supposed to form spin S-T configuration with non-zero energy splitting at zero magnetic field due to exchange interaction, while an odd number of electrons leaves a dangling spin whose two spin states are degenerate at zero magnetic field. These explain why we observe excited states for the even number of electrons and not for the odd numbers. Third, the magnitude of these excited states is comparable with the S-T splitting observed in  pump-and-probe \cite{Kouwenhoven-Spin-ST-T1} and biased transport experiments \cite{Kouwenhoven-Spin-Spectroscopy}. 

\begin{figure}[t]
\begin{center}
\epsfig{file=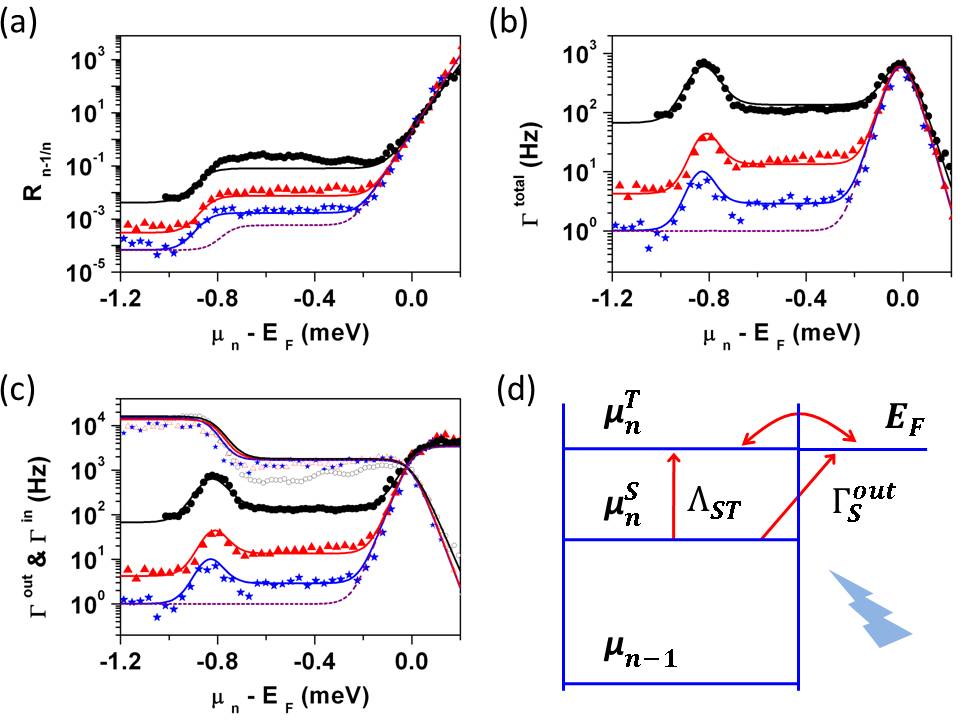, width=1\linewidth, angle=0}
\end{center}
\vspace{-7.5mm}
\caption{ (a) - (c): RTS statistics of $1e \leftrightarrow 2e$ transition under different back-action strength. Back-action from weak to strong: blue stars are taken at $V_{T}=-1.5V$ and $V_{dc}=1mV$; red triangles at  $V_{T}-1.5V$ and $V_{dc}=2mV$; black circles at $V_{T}=-1.2V$ and $V_{dc}=1mV$, same as the data in Fig. \ref{Figure3} for $n=2$. In (c), closed shapes are $\Gamma^{out}$ and open shapes are $\Gamma^{in}$. Solid lines are our simulation. The purple dashed lines are the simulation at the same condition as $V_{T}=-1.5V$ and $V_{dc}=1mV$, except that  $T_{1}$ is reduced from $0.7ms$ to $7ns$. (d) Illustration of the back-action driven excitation between a ground state S and an excited state T.}
\label{Figure4}
\end{figure}

We think that the observed excitation from spin singlet to triplet states must be due to the QPC back-action. People usually use large bias across the QD contacts or fast pulses on the surface gates to pump electrons onto the excited states. In our experiment, no pulse is applied and both the QD contacts are grounded. There seems no other excitation source except the back-action. To verify this assertion, we tuned the back-action down by closing the QD-QPC gap or decreasing the QPC dc bias. Fig. \ref{Figure4} (a) - (c) show the $1e \leftrightarrow 2e$ RTS statistics under different back-action strength. For the black circles, red triangles, and blue stars, the back-action is from strong to weak and we can clearly see that both the saturation and excitation effects become weaker. 

The excitation by back-action is illustrated in Fig. \ref{Figure4} (d). Suppose $ne$ state has a ground level $E^{S}_{n}$ and an excited level $E^{T}_{n}$. When $\mu^{S}_{n} << E_{F}$, the $n^{th}$ electron is supposed to remain on $E^{S}_{n}$, and $E^{T}_{n}$ is kept empty due to Coulomb repulsion. However, the back-action drives the $n^{th}$ electron out of the dot or up to $E^{T}_{n}$, and initiates the transition between $E^{T}_{n}$ and the 2DEG. So we see a side peak when $\mu^{T}_{n}$ is aligned with  $E_{F}$. To quantitatively understand this effect, we developed a phenomenological theory. We consider the general case of $1$ state for $(n-1)e$ and $M$ states for $ne$: $E^{\alpha}_{n}$ ($\alpha=1, ..., M$). Their occupancy can be described by a series of rate equations \cite{Rate-Equation-Theory}:
\begin{center}
$\frac{d}{dt} P_{i}= \sum_{j} (\Gamma_{ji} - \delta_{ji}  \sum_{k} \Gamma_{jk}) P_{j}$
\end{center}

Here subscripts $i, j, k=0, 1, ..., M$. $P_{0}$ denote the $(n-1)e$ occupancy and $P_{\alpha}$ denotes the occupancy for each  $ne$ state $E^{\alpha}_{n}$. $\Gamma_{ji}$ is the tunneling rate from state $j$ to $i$. Specially, $\Gamma_{ii}=0$, $\Gamma_{\alpha0}=\Gamma_{\alpha}^{out}$, and $\Gamma_{0\alpha}=\Gamma_{\alpha}^{in}$. For $\beta>\alpha$, $(\Gamma_{\beta \alpha}-\Gamma_{\alpha \beta})$ is the relaxation rate from $E^{\beta}_{n}$ to $E^{\alpha}_{n}$. Because $P_{0}+ \sum_{\alpha} P_{\alpha}=1$, we can reduce the steady-state rate equations  $\frac{d}{dt} P_{i}=0$ to $M$ independent equations:
\begin{center}
$\sum_{\beta}X_{\alpha \beta} P_{\beta} = Y_{\alpha}$
\end{center}
\begin{center}
$\left\{ \begin{array}{llll}
X_{\alpha \beta}= \Gamma_{0\alpha} - \Gamma_{\beta \alpha} +\delta_{\beta \alpha}(\Gamma_{\beta 0}+\sum_{\gamma}\Gamma_{\beta \gamma}) \\
Y_{\alpha}=\Gamma_{0\alpha} \\
\end{array} \right. $
\end{center}

These equations can be easily solved if all the rates $\Gamma_{ji}$ are known. Then the RTS statistics can be obtained by:
\begin{center}
$\left\{ \begin{array}{llll}
\Gamma^{out}= \sum_{\alpha} (P_{\alpha}\Gamma_{\alpha 0}) / \sum_{\alpha} P_{\alpha} \\
\Gamma^{in}= \sum_{\alpha} \Gamma_{0\alpha} \\
\Gamma^{total}=1/(1/\Gamma^{out}+1/ \Gamma^{in}) \\
R_{n-1/n}=\Gamma^{out}/\Gamma^{in}
\end{array} \right. $
\end{center}

As discussed above, QPC back-action drives the QD electrons out of the dot or up to the excited states. We specify these excitation rates as $\Lambda_{\alpha 0}$ and $\Lambda_{\alpha \beta}$. Since they were found to be energy independent before a cut-off energy, we set all the tunneling rates $\Gamma_{ji}$ as follows \cite{Ming-Submitted-RTS-Backaction}: 
\begin{center}
$\left\{ \begin{array}{llll}
\Gamma_{\alpha 0}=g_{n-1} [\Lambda_{\alpha 0} + \Gamma_{\alpha}^{*}  (1-f(\mu_{n}^{\alpha}))] \\
\Gamma_{0\alpha}=g_{n} \Gamma_{\alpha}^{*}  f(\mu_{n}^{\alpha}) \\
\Gamma_{\alpha \beta}= \Lambda_{\alpha \beta}, \Gamma_{\beta \alpha}= \Lambda_{\alpha \beta}+1/T_{1}^{\beta \alpha} & (\beta>\alpha)
\end{array} \right. $
\end{center}

Here $\Gamma_{\alpha}^{*}$, $\Lambda_{\alpha 0}$, $\Lambda_{\alpha \beta}$, and $T_{1}^{\beta \alpha}$ are input as free parameters to simulate the observed RTS statistics.  $g_{n-1}=2$ is the $(n-1)e$ spin degeneracy, and $g_{n}=1$ since now we consider each $ne$ state separately. For convenience, we assume $\Lambda_{\alpha 0} / \Gamma_{\alpha}^{*}$ is the same for all $\alpha$ and $\Lambda_{\alpha \beta}$  remains constant for all $\alpha \beta$, because the back-action phonon spectrum was revealed to be invariant within a wide energy range \cite{Ming-Submitted-RTS-Backaction}. The comparison with the data at extreme conditions can give us estimate about the values of $\Gamma_{\alpha}^{*}$ and $\Lambda_{\alpha 0}$. For instance, $\Gamma^{in} \approx g_{n} \Gamma_{1}^{*}$ when $\mu_{n}^{1}<<E_{F}<<\mu_{n}^{2}$; $\Gamma^{in} \approx g_{n} \sum_{\alpha} \Gamma_{\alpha}^{*}$ and $\Gamma^{out} \approx g_{n-1} \Lambda_{\alpha 0}$ when  $\mu_{n}^{M}<<E_{F}$; and so on. Some relaxation time $T_{1}^{\beta \alpha}$ are measurable through our pump-and-probe measurements, and we use these values to assist our simulation.

For $n=2$, two spins form a  ground state (singlet $|S\rangle$) and three energy-degenerate excited states (triplet $|T^{-}\rangle$, $|T^{0}\rangle$, and $|T^{+}\rangle$). In principle, we cannot distinguish the three triplet states at zero magnetic field \cite{Backaction-Magnetic-Field}. The observed excited state should be an average effect for all of them. However, their tunneling rate and relaxation time should be different.  Presumably, one of them should dominate due to its overwhelming relaxation time and/or tunneling rate. For instance, pump-and-probe measurement usually only reveals a magnetic field invariant triplet state ($|T^{0}\rangle$) which possesses a long relaxation time \cite{Kouwenhoven-Spin-ST-T1}. In this paper we don't pretend to know which state dominates and just denote it as $|T\rangle$ without losing generality. After setting these rules, we gave the simulation results for the data in Fig. \ref{Figure4} (a) - (c), shown as the solid lines. Our simulation clearly repeats all the major features. As we increase the back-action strength, the tunneling rate $\Gamma_{\alpha}^{*}$ changes very little: $\Gamma_{S}^{*}$ varies between $1.7kHz$ and $1.8kHz$, and $\Gamma_{T}^{*}$ from $12kHz$ to $14kHz$. The simulated relaxation time $T_{1}$ increases from $0.7ms$ through $0.8ms$ to $1.9ms$, probably due to the quantum Zeno effect \cite{Quantum-Zeno-Effect}. What we are most concerned in this experiment is that the back-action induced tunneling rates $\Lambda_{\alpha 0}$ and $\Lambda_{\alpha \beta}$ show dramatic growth: $\Lambda_{S0}=\Lambda_{T0}$ increases from $0.5Hz$ through $2Hz$ to $15Hz$, and $\Lambda_{ST}$  from $2Hz$ through $10Hz$ to $120Hz$. The increase of spin S-T excitation rate $\Lambda_{ST}$ proves that it is driven by back-action.

In addition, we simulated the case of very short relaxation time $T_{1}$. For $V_{T}=-1.5V$ and $V_{dc}=1mV$, we intentionally reduced $T_{1}$ from $0.7ms$ to $7ns$ while keeping all other parameters unchanged, in resemblance of an orbital excited state. The simulation (purple dashed lines) shows that the strong side peak in $\Gamma^{total}$ and $\Gamma^{out}$ at $-0.80meV$ vanishes. Only some residue remains in $R_{n-1/n}$ and $\Gamma^{in}$. This verifies that charge excited states have low visibility due to their short relaxation time.

For $n=4$, we need to consider two spin excited states since we observed two extra features, as indicated by the red arrows in Fig. \ref{Figure3}. First, the simulation reveals that $\Lambda_{\alpha 0}=18Hz$ and $\Lambda_{\alpha \beta}=90Hz$. It is these strong back-action induced tunneling rates that makes the observation of spin excited states possible. To get a better understanding of the excited states spectroscopy, we summarize the results for all the even electron numbers in Table I, where $\Delta E^{\alpha 1}=E_{n}^{\alpha}-E_{n}^{1}$ is the excited state energy with respect to the ground state; $E_{C}$ is the charging energy; and $\Delta \epsilon$ is the estimated orbital level spacing. We concluded that if the ground state ($\alpha=1$) is the spin singlet on the first orbital level ($|1S\rangle$), then the second state ($\alpha=2$) is the spin triplet on the first orbital level ($|1T\rangle$), and the third state ($\alpha=3$) is the spin singlet on the second orbital level ($|2S\rangle$). First, the relaxation time $T_{1}^{21}$ ($|1T\rangle \rightarrow |1S\rangle$) and $T_{1}^{32}$ ($|2S\rangle \rightarrow |1T\rangle$) are long since they are between singlet and triplet states. $T_{1}^{31}$ ($|2S\rangle \rightarrow |1S\rangle$) turns out to be short, since this is a non-spin-flipping process. Second, the ratio of $\Delta E^{21}$ ($|1S\rangle-|1T\rangle$ energy splitting) for $n=6$, $4$ and $2$ is $0.38:0.50:0.80=0.48:0.63:1$, in agreement with that of the charging energy $E_{C}$, $3.5: 4.5:6.6=0.53:0.68:1$. Recall that both the charging energy and exchange energy (thus the S-T splitting) are proportional to the reciprocal of the QD size, this agreement implies that the increase of 1S-1T splitting with decreasing electron number is due to the shrinking of the QD size. Last, $\Delta E^{31}$ ($|1S\rangle-|2S\rangle$ energy splitting) should actually be the orbital level spacing, as proved by their good agreement with our estimated $\Delta \epsilon$. All these agreements validate our energy spectroscopy assignments. Here we need to point out two things: For $n=2$, the energy of the third state $|2S\rangle$ is out of the shown energy window; For $n=6$, the signal-to-noise ratio is too small for us to perform a faithful simulation, probably due to much smaller energy splitting, tunneling rate, and/or relaxation time.
\begin{table}[htb] 
\caption{Simulation results for $n=2$, $4$, and $6$.} 
\label{Table1}
\begin{center}
\begin{tabular}{|c|c|c|c|c|c|c|c|}
\hline                  
$n$ & $T_{1}^{21}$ & $T_{1}^{31}$ & $T_{1}^{32}$ & $\Delta E^{21}$ & $\Delta E^{31}$ &  $E_{C}$ & $\Delta \epsilon$ \\
\hline  
$6$ & $/$ & $/$ & $/$ & $0.38 meV$ & $0.62 meV$ & $3.5 meV$ & $0.59 meV$ \\     
\hline
$4$ & $1.2 ms$ & $25us$ & $1.0 ms$ & $0.50 meV$ & $0.90 meV$ & $4.5 meV$ & $0.98 meV$ \\
\hline
$2$ & $1.9 ms$ & $/$ & $/$ & $0.80 meV$ & $/$ & $6.6 meV$ & $2.12 meV$ \\
\hline
\end{tabular}
\end{center}
\end{table}

The ability to determine the back-action induced tunneling rate from spin ground state to spin excited states enables us to evaluate the influence of back-action on the operation of individual electron spin states, based on which spin qubits are implemented. For instance, when performing the coherent oscillation of a spin S-T based qubit, the averaging of the QPC read-out over a time duration comparable to the relaxation time $T_{1}^{ST}$ is repeated many times \cite{}. If during each read-out cycle, the unwanted back-action induced spin S-T excitation is severe, then the qubit status is ruined (The back-action induced tunneling out rate $\Lambda_{\alpha 0}$ can be suppressed by closing the tunneling barriers so its effect is not discussed here.). So we use the ratio between back-action driven spin S-T excitation rate $\Lambda_{ST}$ and the relaxation rate $1/T_{1}^{ST}$ to set an upper-limit of the fidelity of qubit operation:
\begin{center}
$Fidelity \leq (1-\Lambda_{ST} T_{1}^{ST})$
\end{center}
  
For $n=2$, we found the back-action imposed upper-limit for fidelity reaches $99.8\%$ for $V_{T}=-1.5V$ and $V_{dc}=1mV$, and drops to $77.2\%$ for $V_{T}=-1.2V$ and $V_{dc}=1mV$. For the first case $\Lambda_{ST}$ is only $2Hz$ and for the latter case $\Lambda_{ST}= 120 Hz$ is much higher. We can see that the increase of back-action strength is possible to cause substantial degradation of the qubit operation fidelity. For a given system, our method provides a way to explicitly determine $\Lambda_{ST}$ and quantitatively evaluate the influence of back-action on qubit operation.

In conclusion, due to the strong and tunable back-action in our device, we observed robust signatures of spin S-T excitations through RTS statistics. We developed a method to quantitatively analyze the back-action driven excitation rate, which opens a door to evaluate the effect of back-action on spin S-T based qubits. We hope this study can raise more interests in the role of back-action in quantum computation.

This work was supported by the NFRP 2011CBA00200 and 2011CB921200, NNSF 10934006, 11074243, 10874163, 10804104, 60921091.


\begin{thebibliography}{99}
\bibitem{Backaction-Magnetic-Field} An experiment under magnetic field is undergoing to distinguish these states.
\end{thebibliography}
\end{document}